\documentclass[sigconf]{acmart}

\AtBeginDocument{%
  \providecommand\BibTeX{{%
    \normalfont B\kern-0.5em{\scshape i\kern-0.25em b}\kern-0.8em\TeX}}}

\copyrightyear{2022}
\acmYear{2022}
\setcopyright{rightsretained}
\acmConference[RecSys '22]{Sixteenth ACM Conference on Recommender Systems}{September 18--23, 2022}{Seattle, WA, USA}
\acmBooktitle{Sixteenth ACM Conference on Recommender Systems (RecSys '22), September 18--23, 2022, Seattle, WA, USA}
\acmDOI{10.1145/3523227.3547377}
\acmISBN{978-1-4503-9278-5/22/09}
\acmPrice{15.00}

\begin{document}

\title{Reusable Self-Attention Recommender Systems in Fashion Industry Applications}

\author{Marjan Celikik}
\email{marjan.celikik@zalando.de}
\affiliation{
  \institution{Zalando SE}
  \city{Berlin}
  \country{Germany}
}
\author{Jacek Wasilewski}
\email{jacek.wasilewski@zalando.de}
\affiliation{
  \institution{Zalando SE}
  \city{Berlin}
  \country{Germany}
}
\author{Ana Peleteiro Ramallo}
\email{ana.peleteiro.ramallo@zalando.de}
\affiliation{
  \institution{Zalando SE}
  \city{Berlin}
  \country{Germany}
}

\renewcommand{\shortauthors}{Celikik, et al.}

\begin{abstract}
A large number of empirical studies on applying self-attention models in the domain of recommender systems are based on offline evaluation and metrics computed on standardized datasets. Moreover, many of them do not consider side information such as item and customer metadata although deep-learning recommenders live up to their full potential only when numerous features of heterogeneous type are included. Also, normally the model is used only for a single use case. Due to these shortcomings, even if relevant, previous works are not always representative of their actual effectiveness in real-world industry applications. In this talk, we contribute to bridging this gap by presenting live experimental results demonstrating improvements in user retention of up to 30\%. Moreover, we share our learnings and challenges from building a re-usable and configurable recommender system for various applications from the fashion industry. In particular, we focus on fashion inspiration use-cases, such as outfit ranking, outfit recommendation and real-time personalized outfit generation. 
\end{abstract}



\maketitle

\section{INTRODUCTION}

\begin{figure*}[t]

  \includegraphics[width=0.6\linewidth]{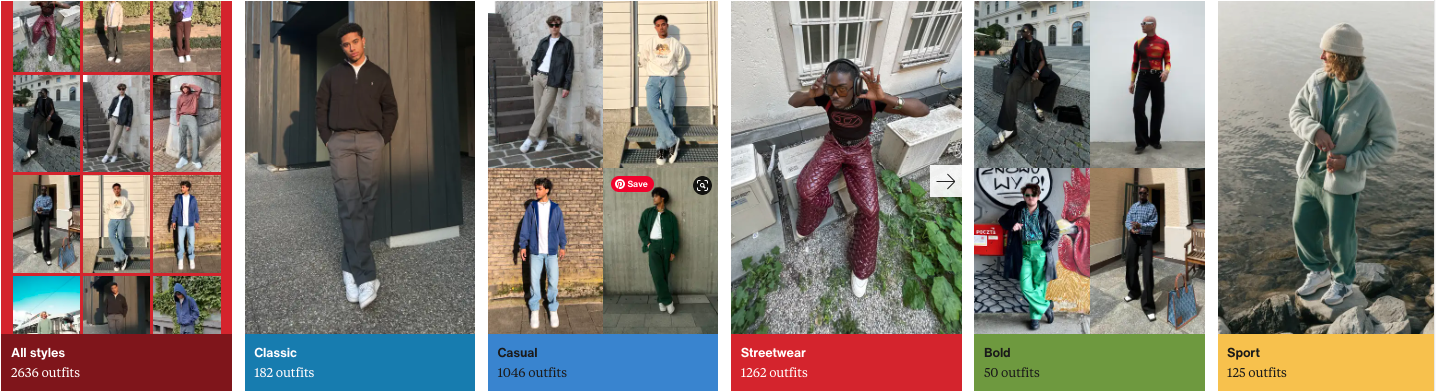}
  \includegraphics[width=0.6\linewidth]{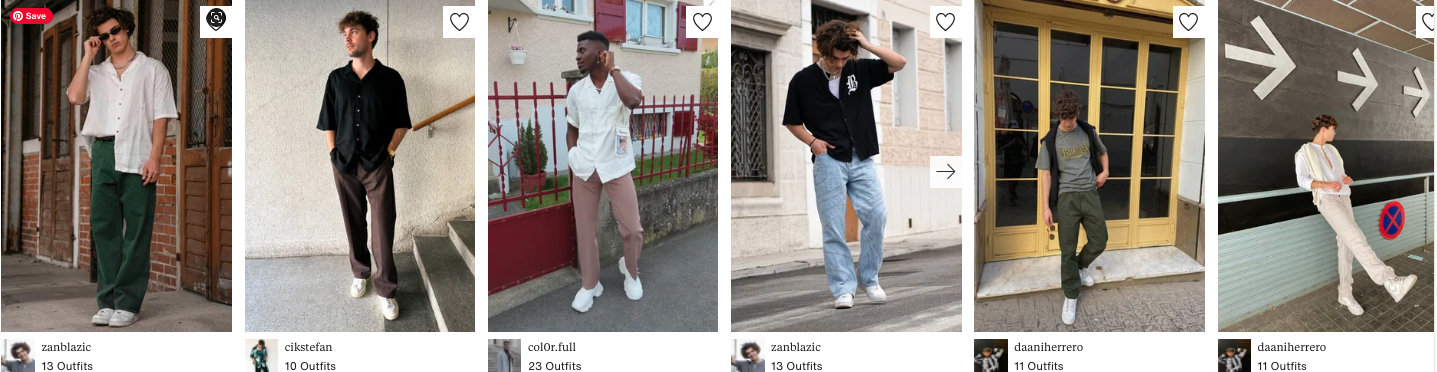}
  \includegraphics[width=0.6\linewidth]{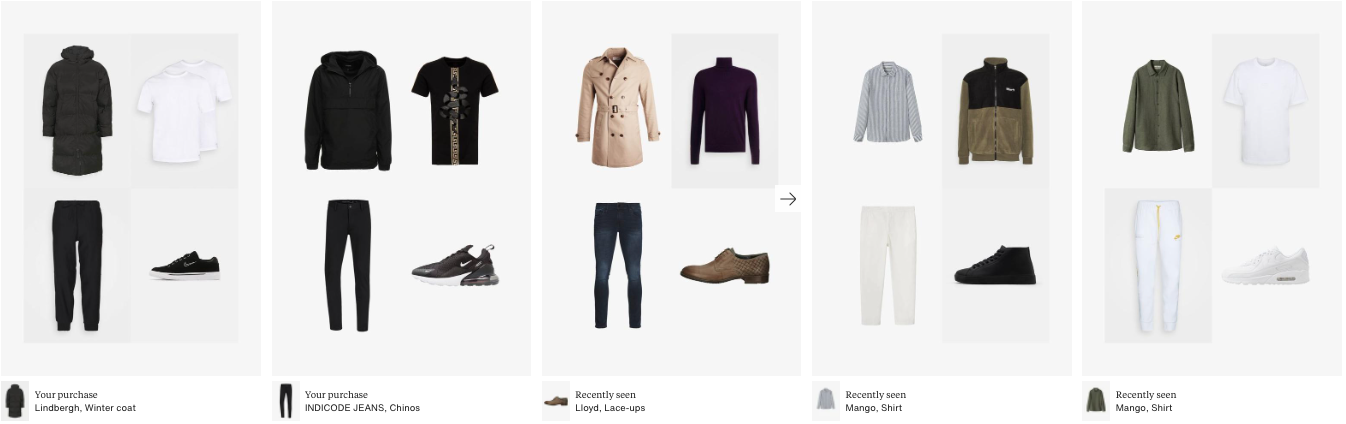}
\caption{"Personalized ranking in the outfit catalog (top-left) and style preview carousel, showcasing outfits in different styles (top-right); "Get-the-look" carousel showing in-session outfit recommendations based on customer's recent interactions (middle); Algorithmic fashion companion (AFC) showcasing algorithmically generated outfits styling recent customer purchases and views (bottom).}
\label{fig:usecases}
\end{figure*}

One of Zalando’s main goals is to become a starting point for fashion inspiration. One way this is achieved is by providing editorial content consisting of creators showcasing fresh and trendy outfit ideas as well as presenting algorithmically generated outfits to style items our customers have purchased, viewed or wishlisted. Besides content quality, the main driver of engagement and customer retention is personalization of the outfit experience on different customer touch points throughout the customer journey, see Figure \ref{fig:usecases}. Historically, all of these customer touch points have been served by different end-to-end systems suited for different use-cases (e.g. ranking and recommendations for long-term interests, in-session recommendations, recommendations for related items, etc.), resulting in increased complexity and significant maintenance cost. Another complexity in our setting is that the customers can interact not only with outfits but also with other fashion entities along the customer journey such as fashion items or creators. In addition, each of these interactions might carry a different signal (e.g. click, wishlist, or purchase). Finally, as only a subset of all customers on the platform interact with outfits, we are also dealing with the cold-start problem. 

\section{MODEL ARCHITECTURE}

In contrast to the common belief that different recommenders are suited for different use-cases \cite{netflixdl}, we showcase that a single Transformer-based recommender system can be trained on diverse types of interactions coming from various sources and re-used across many related use-cases, such as session-based recommendation for short-term interests as well as personalized ranking based on long-term user preferences. This has two benefits, first, it significantly increases the training dataset size, and second, it dampens feedback loops, where the recommender system is trained on data from the same carousel on a previous day. Feedback loops can amplify biases such as popularity and presentation bias \cite{netflixdl,Chaney2018,criticialfashionxrecsys}. The same model is used to serve partial cold-start users by utilizing other types of interactions in the session as well as full cold-start users without interactions by utilizing “static” contextual information about the user.

We model each user as a sequence of interactions, where each interaction can be performed with a different entity (e.g. item, outfit or a creator) and be of different type (e.g. click, purchase, wishlisting). We train on every item in the sequence, but predict only on one of entities, e.g. outfits, and only those contribute to the loss thanks to the introduced target boolean mask. For our \textbf{recommendation use-cases}, we use a standard Transformer encoder \cite{sasrec,attention,trm4rec} trained with casual language model (CLM) approach, although the approach is oblivious of the training logic and it can be trained with the masked language modeling (MLM) as well \cite{bert4rec,trm4rec}. For the \textbf{personalized outfit generation use-cases}, we use variations of the standard encoder-decoder (sequence-to-sequence) Transformer architecture \cite{attention} as in \cite{alibabapog,amazonoutfit}. The model is trained on the same source of sequence interactions. However, since one of the main task of the latter model is to learn fashion compatibility among different items, the targets that are fed into the decoder are the items in the interacted outfits, while the preceding interactions to the outfit interaction are fed into the encoder.

\begin{figure*}[t]
  \centering
  \includegraphics[width=\linewidth]{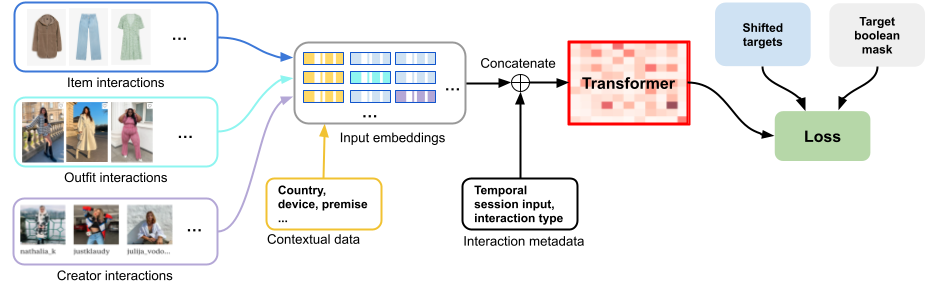}
  \caption{Overview of the model trained on different interaction entities and types that are converted into embeddings with the same structure. Temporal session and interaction data is concatenated directly to the learned embeddings. Masking is used in order to consider target outputs only of certain entity type (e.g. outfits).}
  \label{fig:model}
\end{figure*}

\subsection{Input Embeddings}

Each entity in the input sequence is represented as a concatenation of its heterogeneous inputs, for example, learned embeddings that correspond to categorical features. Missing features are padded with 0s, e.g. an outfit may have a creator while an item does not. Since an outfit is a set of items, we use a simple representation where we average the embeddings from the same categorical feature coming from different items. Individual items in the sequence are represented as single-item outfits and creators are treated as set of outfits. We concatenate the 1-hot encoding of interaction features to the entity embedding. The user context is encoded as an embedding with the same dimensionality as the item embeddings and set at the first position(s) of the Transformer. 

\subsection{Modeling Sessions for Long and Short-Term Interests}

An interaction sequence consists of different sessions, where a session is a list of interactions within a given time frame (e.g., a day) when the user has a clear intent while their interests can change dramatically over different sessions. Modeling the sequences directly while ignoring this structure affects performance negatively, which we also observed in our experiments. We model sessions by introducing temporal inputs in the form of interaction recency, defined as the number of days between the action timestamp and model training/serving timestamp. We discretize recency and consider only the integer part of the timestamp. Each interaction is assigned its own recency that is concatenated with the rest of the item features. Hence, the attention mechanism can select the interactions in the sequence that are most relevant for the prediction. Figure \ref{fig:model} provides a summary of the modeling choices and the different sources of data used for training.

\section{LIVE EXPERIMENTS}

Table \ref{tab:reco} and Table \ref{tab:afc} show live experiments comparing our recommendation and outfit generation algorithms to various existing algorithms (LTR \cite{TensorflowRankingKDD2019}, CNN-embedding-based kNN, Siamese Nets \cite{siamesenets}) on different premises as illustrated in Figure \ref{fig:usecases}. Our new algorithms outperforms all of the baselines by large margins.

\begin{table*}[h]
  \caption{A/B test results showing increase in user engagement (and retention) of our Transformer-based recommender compared to the existing algorithms on different premises illustrated in Figure \ref{fig:usecases}.}
\begin{tabular}{cccccc}
\toprule
\textbf{Segment} & \textbf{Outfit catalog (LTR)} & \textbf{Style preview carousel (LTR)} & \textbf{Get-the-look (CNN-kNN)} \\
\midrule
 \textbf{Cold-start}      & +42.0\%            & +39.7\%     & +109.4\%  \\
 \textbf{Existing}        & +27.0\%            & +23.1\%     & +130.6\%  \\      
 \textbf{All}             & +28.5\%            & +23.9\%     & +130.4\%  \\    
\bottomrule
\end{tabular}
\label{tab:reco}
\end{table*}

\begin{table*}[h] 
  \caption{A/B test results on AFC (Figure \ref{fig:usecases}, bottom) comparing user engagement between (1) the Transformer algorithm for personalized outfit generation vs. the existing Siamese Nets \cite{siamesenets}. (2) The same Transformer-based algorithm with a fallback model (in order to deal with undersupply due to business rule filtering): a more efficient version of the Transformer that predicts a single embedding in its output layer vs. Siamese Nets.}
\begin{tabular}{ccc}
\toprule
        & \textbf{(1) Transformer vs. SN} & \textbf{(2) Transformer + Transformer w/ Embed. vs. Transformer + SN} \\
\midrule
\textbf{Engagement} & +5.79\%  &  +1.54\%  \\
\textbf{Retention}  & +11.20\% &  +4.9\%   \\
\bottomrule
\end{tabular}
\label{tab:afc}
\end{table*}

\section{PRACTICAL DESIGN CHOICES AND SCALABILITY}

We designed a flexible and configurable recommender system that can be re-used throughout different use-cases thanks to configurable filtering logic embedded in the backend. Our design proved to be particularly practical and scalable and it follows \textbf{inputs $\rightarrow$ scoring $\rightarrow$ filtering $\rightarrow$ re-ranking} framework. Our system works with real-time as well as with daily provided interaction data. The items are scored by feeding the final output of the last position of the Transformer into a softmax layer. Other ranking functions such as bpr, top1 \cite{rnns} and binary-cross entropy \cite{sasrec} proved less effective in our setting. The filtering and the re-ranking logic is applied per use-case. The re-ranking phase consists of applying freshness rules and diversification. Examples for freshness rules include re-ranking by including item age feature toggle, exponential age decay, etc. In our case the latter proved particularly effective since it did not harm relevance and substantially increased fresh content among the top-k recommendations. Examples for diversifying the result list include diversification heuristics based on similarity and balancing exploration by introducing sampling via the multi-armed bandit framework.

In a typical microservice backend architecture, the data required to be passed to the serving endpoint needs to be fetched from an in-memory database and transformed into features that should be fed into the model by the same logic used during training. Since the interaction data can be an outfit or a creator which are all regarded as sets of items, the number of calls as well as the amount of data needed to be send over the wire increases substantially in our setting. This is aggravated in the outfit generation scenario where due to the auto-regressive nature of outfit generation process, the backend service has to wait for the next item in the outfit. This significantly increases the complexity in the backend as well as the network overhead. Therefore, in order to avoid duplication of the serving and training logic, avoid high network latencies and decrease the backend complexity, we couple the input data processing with the actual model. This means that the actual entity data lives as a mapping on the same service as the model and hence the same logic is used for training and inference. Note that this approach did not cause missing or data staleness issues in our setting since we retrain our model and update the entity data on a daily basis. The serving endpoint is fed only with raw interaction ids. Thanks to these design choices, our system had roughly twice as low p99 latency and scaled substantially better (roughly by a factor of 3 for the same amount of traffic) compared to the existing ranking system based on TensorFlow-ranking \cite{TensorflowRankingKDD2019} that due to scalability issues re-ranks only 1000 outfits per request.

\section{SPEAKER BIO}

\textbf{Marjan Celikik} is a principal applied scientist at Zalando working on problems related to personalization, recommender systems and productionizing large-scale machine learning projects. He holds a PhD in Information Retrieval and Search Engines from University of Freiburg and the Max-Planck-Institute for Computer Science and a Master's in Computer Science from University of Saarland.

\bibliographystyle{ACM-Reference-Format}
\bibliography{industry_talk_zalando}

\end{document}